\documentclass[a4paper,twoside]{article}

\usepackage{epsfig}
\usepackage{subfigure}
\usepackage{calc}
\usepackage{amssymb}
\usepackage{amstext}
\usepackage{amsmath}
\usepackage{amsthm}
\usepackage{multicol}
\usepackage{pslatex}
\usepackage{apalike}

\usepackage{graphicx}
\usepackage{float}
\usepackage{color}
\definecolor{mygreen}{RGB}{28,172,0} 
\definecolor{mylilas}{RGB}{170,55,241}
\usepackage{tikz}
\usetikzlibrary{calc,matrix}
\usepackage{sgame}
\usepackage{fancyhdr}

\usepackage[utf8]{inputenc}
\usepackage{bm}
\usepackage[english]{babel}
\usepackage[toc,page]{appendix}
\usepackage{listings}
\usepackage{caption}
\usepackage[toc,page]{appendix}

\theoremstyle{plain}
\newtheorem{thm}{Theorem}[section]

\newtheorem{prop}[thm]{Proposition}

\theoremstyle{definition}
\newtheorem{defn}{Definition}[section]

\newcommand*{\Scale}[2][4]{\scalebox{#1}{$#2$}}
\usepackage{SCITEPRESS}     

\begin{document}

\title{Distributionally Robust Games with Risk-Averse Players} 

\author{\authorname{Nicolas Loizou}
\affiliation{School of Mathematics, University of Edinburgh, Edinburgh ,U.K.}
\email{}
}

\abstract{ We present a new model of incomplete information games without private information in which the players use a distributionally robust optimization approach to cope with the payoff uncertainty. With some specific restrictions, we show that our ``Distributionally Robust Game" constitutes a true generalization of three popular finite games. These are the Complete Information Games, Bayesian Games and Robust Games. Subsequently, we prove that the set of equilibria of an arbitrary distributionally robust game with specified ambiguity set can be computed as the component-wise projection of the solution set of a multi-linear system of equations and inequalities. For special cases of such games we show equivalence to complete information finite games (Nash Games) with the same number of players and same action spaces. Thus, when our game falls within these special cases one can simply solve the corresponding Nash Game. Finally, we demonstrate the applicability of our new model of games and highlight its importance.}

\onecolumn \maketitle \normalsize \vfill

\section{\uppercase{Introduction}}
\label{sec:introduction}

\noindent
The classical, complete-information finite games assume that the problem data (in particular the payoff matrix) is known exactly by all players. In a now famous result(\cite{nash1950equilibrium}, \cite{nash1951non}), Nash has shown that any such game has an equilibrium in mixed strategies. More specifically, in his formulation Nash assumed that all players are rational and that all parameters(including payoff functions) of the game are common knowledge. With these two assumptions, the players can predict the outcome of the game. For this reason each player given the other players strategies is in a position to choose the mixed strategy that gives him the maximum profit. A tuple of these strategies is what we call ``Nash Equilibrium". 

The existence of equilibria in mixed strategies was later extended to a class of incomplete information finite games by Harsanyi \cite{harsanyi2004games}, who assumed that the payoff matrix is not known exactly but rather represents a random variable that is governed by a probability distribution known to all players. In particular, Harsanyi assumed that a full prior distributional information for all parameters of the game is available and that all players use this information in order to compute the payoff functions of the game. This computation is made using the Bayes' rule. For this reason these games are called ``Bayesian Games" and their equilibrium ``Bayesian Nash Equilibrium".

In 2006, Aghassi and Bertsimas \cite{aghassi2006robust} proposed a new class of distribution-free finite games where the payoff matrix is only known to belong to a given uncertainty set. This model relaxes the distributional assumptions of Harsanyi's Bayesian games, and it gives rise to an alternative distribution-free equilibrium concept. Furthermore, in this model of games the players use a robust optimization approach to the uncertainty and this is assumed to be a common knowledge. That is, given the other players strategies each player tries to maximise his worst case expected payoff (worst case is taken with respect to the uncertainty set). The using of the robust optimization approach is the reason for calling these ``Robust Games" and their equilibrium ``Robust Optimization Equilibrium".

More recently, Qu and Goh \cite{qu2012distributionally} proposed a distributional robust version of the finite game where they only consider the case in which the players are risk neutral and they focus on an application of supply chain. This model later was extended to continuous games from Sun and Xu \cite{sun2015convergence}. The approach of these papers is in contrast to our framework of modelling the players as each seeking to minimize his worst case CVAR expected loss. Moreover they offer no ideas on computation of their equilibria.

In this paper we present a new model of incomplete information games without private information in which the players use a distributionally robust optimization approach to cope with payoff uncertainty. In our model players only have partial information about the probability distribution of the uncertain payoff matrix. This information is expressed through a commonly known ambiguity set of all distributions that are consistent with the known distributional properties. Similar to the robust games framework, players in distributionally robust games adopt a worst case approach. Only now the worst case is computed over all probability distributions within the ambiguity set. More specifically we use a worst case CVaR(Conditional Value at Risk) approach. This allows players to have several risk attitudes which make our model even more coveted since in real life applications players rarely are risk neutral.

For classical work on slightly related game models we refer the reader to \cite{hayashi2005robust} and \cite{nishimura2012semidefinite}. The recent paper \cite{singhexistence} also deals with similar model of games but in contrast to our approach the authors focused on the existence of equilibrium.

The remainder of this work is organized as follows: Section 2 introduces our notation as well as definitions that are used through the paper. Section 3 proposes and analyses our new model for Distributionally Robust Games. After formulating the model, we show that any other finite game can be expressed as a distributionally robust game. In Section 4, we prove the equivalence of the set of equilibria of a distributionally robust game and the component-wise projection of the solution set of multi-linear system of equations and inequalities. Section 5 shows the equivalence of distributional robust games and Nash games in special cases. Sections 6 and 7 are devoted to an illustrative example and numerical experiment, respectively. Finally, conclusions and future directions are drawn in Section 8.

All proofs are relegated to the appendix.

\section{\uppercase{Notations-Definitions}}
\noindent The following notational conventions are used in this paper. Boldface upper case letters will denote matrices and boldface lower case letters will denote vectors. To denote uncertainty we will use the  tilde ($\tilde{\cdot}$) in which the input parameter ($\cdot$) can be either scalar, vector, or matrix and the check ($\check{\cdot}$) will indicate the nominal counterpart of the uncertain coefficient $\tilde{\cdot}$. Finally, $vec(\bm{A})$ denotes the column vector obtained by stacking the row vectors of the matrix $\bm{A}$ one on top of the other.\\
With $\bm{\check{P}} \in \mathbb{R}^{N\times\prod\limits_{i=1}^N a_i} $ we denote the payoff matrix of a complete information game (fixed matrix)  while with $\bm{\tilde{P}}$ the uncertainty matrix of the incomplete information games.\\
In particular,  ${\bm{\check{P}}}_{(j_1,j_2,....j_N)}^i$ denotes the payoff to player $i$ when player $k \in \{1,2,....N\}$ plays action $j_k \in \{1,2,....,a_k\}$ and 
$S_{a_i} = \{ \bm{x^i} \in {\mathbb{R}}^{a_i}| \bm{x^i} \geq 0, \sum\limits_{{J_i}=1}^{a_i} {x^i}_{j_i}=1 \}$ expresses the set of all possible mixed strategies of player $i$ over all actions $\{1,2,...a_i\}$. Moreover, let $\pi_i(\bm{P;x^1,x^2,...x^N})$ indicate the expected payoff of player i when the payoff matrix is given by $\bm{P}$ and player $k \in \{1,2,....N\}$ plays mixed strategy $\bm{x^k} \in S_{a_k}$. That is,
\begin{equation}
 \Scale[0.9] { \pi_i(\bm{P;x^1,x^2,}...\bm{x^N}) = \sum\limits_{{j_1}=1}^{a_1}...\sum\limits_{{j_i}=1}^{a_i}... \sum\limits_{{j_{N}}=1}^{a_{N}}{\bm{P}}_{(j_1,j_2,....j_N)}^i \prod\limits_{i=1}^N {x^i}_{j_i}}
 \end{equation}
Finally, in this paper, we use exactly like Bertsimas and Aghassi \cite{aghassi2006robust}  the following shorthands:
$$ \bm{x^{-i}} = (\bm{x^1,x^2,..,x^{i-1},x^{i+1},...x^N})$$
$$ (\bm{x^{-i},u^i}) =(\bm{x^1,x^2,..,x^{i-1},u^i,x^{i+1},...x^N})$$
$$ S= \prod\limits_{i=1}^N S_{a_i}  \quad , \quad S_{-i}= \prod\limits_{k=1,k\neq i}^N S_{a_k}$$

\noindent and the following definitions for the equilibrium in Nash games, Bayesian Games and Robust Games:\\

\noindent The tuple of strategies $(\bm{x^1,x^2,...x^N})\in S$ is: 
\begin{itemize}
\item \textbf{Nash Equilibrium} iff for each player $i\in \{1,2....N\}$:
\begin{equation}
\label{lllll}
\bm{x_i} \in \underset{\bm{u^i}\in S_{a_i}}{\operatorname{argmax}}\,\pi_i(\bm{\check{P};x^{-i},u_i})
\end{equation}
\item \textbf{Bayesian Nash Equilibrium} iff for each player $i\in \{1,2....N\}$:
\begin{equation}
\label{lmljsd}
\bm{x_i} \in \underset{\bm{u^i}\in S_{a_i}}{\operatorname{argmax}}\,[\underset{\tilde{P}}E\pi_i(\bm{\tilde{P};x^{-i},u_i)}]
\end{equation}
\item \textbf{Robust Optimization Equilibrium} iff for each player  $i\in \{1,2....N\}$,
\begin{equation}
\label{loioi}
\bm{x_i} \in \underset{\bm{u^i}\in S_{a_i}}{\operatorname{argmax}}\,[\underset{\bm{\tilde{P}}\in U}\inf\pi_i(\bm{\tilde{P};x^{-i},u_i})]
\end{equation}
\end{itemize}
\section{THE NEW MODEL}
\noindent In this section we present the new model of incomplete information games without private information in which the players use a distributionally robust optimization approach to cope with payoff uncertainty. 
We also show that under specific assumptions about the ambiguity set and the values of risk levels, Distributionally Robust Game constitutes a true generalization of Nash, Bayesian and Robust Games.

We introduce the new model of Distributionally Robust Games by first giving the two important definitions of Best Response and Distributionally Robust Optimization Equilibrium and explain them later in details.

\begin{defn}
\label{bestreD}
In the distributionally robust model, for the case without private information, players i's \emph{best response} to the other players strategies $\bm{x^{-i}} \in S_{-i}$ must belong to:
\begin{equation} 
\underset{u^i\in S_{a_i}}{\operatorname{argmin}}\,\underset{Q \in \mathcal{F}}{\text{sup}}\; Q\text{-CVaR}_{\varepsilon_i} [-\pi_i(\bm{\tilde{P};x^{-i},u^i})]  
\end{equation}
\end{defn}

\begin{defn}
\label{DROE}
$(\bm{x^1,x^2,}...\bm{x^N}) \in S$ is said to be a \emph{Distributionally Robust Optimization Equilibrium} of the corresponding game with incomplete information iff $\forall i \in \{1,2,..N\},$
\begin{equation} 
\bm{x^i} \in \underset{u^i\in S_{a_i}}{\operatorname{argmin}}\,\underset{Q \in \mathcal{F}}{\text{sup}}\; Q\text{-CVaR}_{\varepsilon_i} [-\pi_i(\bm{\tilde{P};x^{-i},u^i})]  
\end{equation}
\end{defn}

 Our approach can be considered a concept closely related to both Harsanyi's Bayesian Games \cite{harsanyi2004games} and Robust Games \cite{aghassi2006robust}. 
In more detail, in Bayesian Games we assume that all players of the game know the exact distribution of the payoff matrix. Now, in the Distributionally Robust approach the players do not know the exact distribution. Instead, they are only aware of a commonly known ambiguity set $\mathcal{F}$ of all possible probability distributions $Q$ that satisfy some specific properties. These distributions have no restriction in their form. That is, the ambiguity set may consists of both, discrete and continuous distributions of the payoff matrix.
In addition, similar to the robust games framework, we assume that each player  adopts a worst case approach to the uncertainty. Only now the worst case is computed over all probability distributions within the set  $\mathcal{F}$. For this formulation which is similar to distributionally robust optimization concept we named these games \textbf{Distributionally Robust Games(DRG)} and we refer to their equilibria as \textbf{Distributionally Robust Optimization Equilibria (DROE)}.

\subsection{CVAR and Main Assumptions}
\begin{defn}(Conditional Value at Risk)\\
$\text{CVaR}_\varepsilon$  of a loss distribution $L$ is the expected value of all losses that exceed $(1 - \varepsilon)$-quantile of the distribution. This can be formalized as:
\begin{equation} 
Q\text{-CVaR}_\varepsilon (L) = \min_{\zeta \in \mathbb{R}} \; \zeta + \frac{1}{\varepsilon} \mathbb{E}_Q [ L - \zeta]^+ 
\end{equation}
where $[x]^+ = \max\{x,0\}$.
\end{defn}
Conditional Value at Risk(CVaR) is one of the most popular quantile-based risk measures because of its desirable computational properties \cite{rockafellar2000optimization},\cite{artzner2002coherent}.
Exactly for these properties we chose to introduce CVaR in the formulation of the new model.\\
Using $Q\text{-CVaR}_{\varepsilon_i}$, we allow the players to have several risk attitudes which is a major difference compared to all other finite games (Nash, Bayesian and Robust Games) in which the players are always risk neutral. Important hypothesis is that risk attitude is a fixed characteristic of each player and it cannot be changed depending the game. It is not a notion like the mixed strategy that a player can choose in order to achieve his best response and minimise his loss. More specifically, the parameter $\varepsilon_i \in (0,1)$ determines the risk-aversion of each decision-maker. In detail, if player i has risk level $\varepsilon_i = 1$ this means that he is risk neutral since the Conditional Value at Risk is equal to the expected value of his loss function ($Q\text{-CVaR}_{\varepsilon_i} = \mathbb{E}_Q $). On the other hand if $\varepsilon_i \leq 1$ the player is risk averse and as $\varepsilon_i \longrightarrow 0$ the risk aversion of the player becomes larger. \\
Conclusively, as parameter $\varepsilon_i$ decreases, the value of  $Q\text{-CVaR}_{\varepsilon_i}$ increases and the risk aversion of the player becomes larger and vice versa. \\
With the introduction of a risk measure in our model we take into account not only that each player wishes to maximize his gain (minimize loss) but also how much she/he is willing to risk to achieve this maximum value (minimum value).\\

\noindent \textbf{$1^{st}$ assumption of the new model:}
\emph{ The risk attitude of each player is assumed to be common knowledge. That is, each player knows how much risk averse are the other players and that all other players know that he knows.} \\

In general, CVaR can be calculated from either the probability distribution of gains or the probability distribution of losses. In this paper we decide to follow the original formulation of Rockafellar and Uryasev (see \cite{rockafellar2002conditional} and \cite{rockafellar2000optimization}) and calculate CVAR from the distribution of losses.\\
For this reason, in definitions ~\eqref{bestreD} and ~\eqref{DROE} we use the \emph{expected loss function} of player $i$, $-\pi_i(\bm{P;x^{-i},x^i})$. Loss distributions are also responsible for the use of $ \underset{\bm{u^i}\in S_{a_i}}{\operatorname{argmin}}\,\underset{Q \in \mathcal{F}}{\text{sup}}\;$ instead of $\underset{\bm{u^i}\in S_{a_i}}{\operatorname{argmax}}\,\underset{Q \in \mathcal{F}}{\text{inf}}\; \;$ that we use in robust games. \\
Finally, in the formulation of the DRG we make two more assumptions.\\

\noindent \textbf{$2^{nd}$ assumption:}
\emph{The players commonly know the ambiguity set of all possible distributions (discrete and continuous) of the payoff matrix.} \\

\noindent \textbf{$3^{rd}$ assumption:}
\emph{Each player adopts, like Robust games, a worst case approach to the uncertainty, only now the worst case is computed over all probability distributions within the set  $\mathcal{F}$. In particular we assume that all players use a worst case CVaR approach.}
\subsection{Generalization of All Other Games}
From the formulation of the DRG (definitions of best response and equilibrium) we can easily understand that only two are the parameters that are amenable to change. These are, risk level  $\varepsilon_i$ of each player i and the ambiguity set $\mathcal{F}$.
In particular, we assume that parameter $\varepsilon_i$ which shows the risk level of each player can take any value in the interval $(0,1)$ and we make no assumptions for the ambiguity set. That is, depending on the game that one faces she/he can choose the more suitable properties that the distributions of the ambiguity set must satisfy.
Hence, if we assume some extra constraints for parameter $\varepsilon_i$ and set $\mathcal{F}$, the previous general formulation can become very specific.

Let's assume that all players $i \in \{1,2,...N\}$ have the same risk level, $\varepsilon_i = 1 \; \forall i \in \{1,2,....,N\}$. Then from the definition of CVaR we obtain the following $ \forall i \in \{1,2,....,N\}$:
\begin{equation}
\label{elnet3}
\Scale[0.7] {\begin{aligned}
Q\text{-CVaR}_{\varepsilon_i} [-\pi_i(\bm{\tilde{P};x^{-i},u^i})] 
& = Q\text{-CVaR}_{1} [-\pi_i(\bm{\tilde{P};x^{-i},u^i})] \\
& = \mathbb{E}_Q [ -\pi_i(\bm{\tilde{P};x^{-i},u^i})]  
\end{aligned}}
\end{equation}
\noindent Therefore the definition of best response~\eqref{bestreD} becomes:
\begin{equation}
\label{elnet4}
\Scale[0.7] {\begin{aligned}
\underset{u^i\in S_{a_i}}{\operatorname{argmin}}\,\underset{Q \in \mathcal{F}}{\text{sup}}\; Q\text{-CVaR}_{\varepsilon_i} [-\pi_i(\bm{\tilde{P};x^{-i},u^i})] 
& = \underset{u^i\in S_{a_i}} {\operatorname{argmin}} \,\underset{Q \in \mathcal{F}}{\text{sup}}\; \mathbb{E}_Q [-\pi_i(\bm{\tilde{P};x^{-i},u^i})] \\
& = \underset{u^i\in S_{a_i}} {\operatorname{argmin}} \,\underset{Q \in \mathcal{F}}{\text{sup}}\; [-\pi_i(\bm{\mathbb{E}_Q [\tilde{P}];x^{-i},u^i})]\\
& =  \underset{u^i\in S_{a_i}} {\operatorname{argmax}} \,\inf_{Q \in \mathcal{F}} \; [\pi_i(\bm{\mathbb{E}_Q [\tilde{P}];x^{-i},u^i})] 
\end{aligned}}
\end{equation}
The equality in the second line in the above expression follows from the linearity of expectation operator and the linearity of $\pi_i$: $$\mathbb{E}_Q [\pi_i(\bm{\tilde{P};x^{-i},u^i})]  = [\pi_i(\bm{\mathbb{E}_Q [\tilde{P}];x^{-i},u^i})]$$ where $\bm{\mathbb{E}_Q [\tilde{P}]} $is the component-wise expected value of $\bm{\tilde{P}}$.
The equality in the third line is due to the following properties of linear functions:
$$\max f(x) = - \min [-f(x)]$$ and
$$z\in \underset{x \in S_{a_i}} {\operatorname{argmin}} f(x) = z\in \underset{x\in S_{a_i}} {\operatorname{argmax}} -f(x).$$
Now, using the assumption that $\varepsilon_i = 1 \quad \forall i \in \{1,2,...N\}$ and by choosing the right properties for the probability distributions of the payoff matrix we can specify our formulation and create the desire games.
\begin{thm}
  \label{allequilibria}
  In the DRG, players i's best response to the other players strategies $\bm{x^{-i}} \in S_{-i}$ must belong to:\\
\begin{equation} 
\Scale[0.9] {\underset{u^i\in S_{a_i}}{\operatorname{argmin}}\,\underset{Q \in \mathcal{F}}{\text{sup}}\; Q\text{-CVaR}_{\varepsilon_i} [-\pi_i(\bm{\tilde{P};x^{-i},u^i})] } 
\end{equation}

\noindent If we assume that $\varepsilon_i = 1 \quad\forall i \in \{1,2,...N\}$, then:
\begin{enumerate}
\item If $ \mathcal{F} = \{ Q : \mathbb{E}_Q [\bm{\tilde{P}}] = \bm{\Psi} \}$ the set of DROE is equivalent to that of a classical Nash Game.
\item If the ambiguity set is singleton, that is, $ \mathcal{F} = \{ Q \}$, the DRG have the same equilibria with a related finite Bayesian Game.
\item If $ \mathcal{F} = \{ Q : Q[\bm{W} \cdot vec(\mathbb{E}_Q [\bm{\tilde{P}}])\leq \bm{h}] = 1\}$ the set of DROE is equivalent to that of the classical Robust Game.
\end{enumerate}
\end{thm}  
\section{\uppercase{Computing Sample equilibria of DRG}}
\noindent To examine if one specific tuple of strategies is equilibrium is a typically elementary procedure in any kind of game. However, to find the set of all equilibria of a game, with complete or incomplete information, is a very difficult task. \\
In this section, we present Theorem ~\eqref{alldistribequilibria}, in which we show that for any finite distributionally robust game with a specified ambiguity set and with no private information the set of equilibria is a projection of the solution set of multilinear system of equalities and inequalities. The projection is like in \cite{aghassi2006robust}, the component wise one into a lower dimensional space.
\begin{thm}
\emph{(Computation of Equilibria in Distributionally Robust Finite Games)}\\
  \label{alldistribequilibria}
  Consider the N-player distributionally robust game in which $i \in \{1,2,...,N\}$ has action set $\{1,2,...,a_i\}, \, 1 < a_i < \infty, $ in which the ambiguity set is:
  \begin{equation}
  \label{knkn}
\Scale[0.7] { \mathcal{F} = \{ Q : Q[\bm{\tilde{P}} \in \mathcal{U} ] = 1 , \;\; \mathbb{E}_Q [vec \bm{\tilde{P}}] = \bm{m}, \;\; \mathbb{E}_Q [\left \lVert vec (\bm{\tilde{P}}) - \bm{m} \right \rVert_1] \leq s \}}
\end{equation}
where $\Scale[0.8] { \mathcal{U} = \{ \bm{P} : \bm{W} \cdot vec(\bm{P})\leq \bm{h}\}}$ is bounded and polyhedral set,
and in which there is no private information. The following two conditions are equivalent.\\
\noindent \textbf{Condition 1)} $\Scale[0.7] { (\bm{x^1,x^2,....x^N})}$ is an equilibrium of the distributionally robust game.\\
\textbf{Condition 2)} For all $\Scale[0.8] { i \in \{1,2,....N\}}$ there exists \\
$\Scale[0.8] { \; \alpha_i, \zeta_i, \rho_i \in \mathbb{R}, \quad\gamma_i \in \mathbb{R_+},\quad}$ $ \Scale[0.8] {\bm{\xi^i, \theta^i} \in \mathbb{R}^m} $ and\\  $\Scale[0.8] {\bm{\beta^i,\lambda^i, \kappa^i,\delta^i, \nu^i, \tau^i, f^i, \phi^i, g^i} \in \mathbb{R}^{N \prod_{i=1}^N a_1}, \;} $ 
such that\\
$\Scale[0.8] {(\bm{x^1,x^2,....x^N},\alpha_i, \zeta_i, \rho_i, \gamma_i, \bm{\beta^i,\lambda^i, \kappa^i,\delta^i, \nu^i, \tau^i, f^i, \phi^i, g^i, \xi^i, \theta^i})}$
satisfies:
\begin{equation}
 \label{gsdf}
 \Scale[0.7] {\begin{array}{l@{\qquad}l}
 \quad \zeta_i +  \frac{1}{\varepsilon_i}\alpha_i + \frac{1}{\varepsilon_i}\bm{m}^\top \bm{\beta^i} + \frac{1}{\varepsilon_i}s\gamma = \rho_i, & \bm{e}^\top \bm{x^i}=1,\\
 \quad \alpha_i - \bm{m^\top \lambda^i} +\bm{m}^\top \bm{\kappa^i} + \bm{h^\top \xi^i} \geq 0, & \rho_i\bm{e}^\top \leq \bm{f^\top Y^i(x^{-i})} \\
 \quad \bm{\lambda^i} +\bm{\kappa^i} - \gamma_i\bm{e} \leq \bm{0}, & \bm{\delta^i} +\bm{\nu^i} - \gamma_i\bm{e} \leq \bm{0}\\
 \quad \alpha_i - \bm{m^\top \delta^i} +\bm{m}^\top \bm{\nu^i} + \bm{h^\top \theta^i}+\zeta_i \geq 0 \\
 \quad -\bm{\delta^i} +\bm{\nu^i} + \bm{W^\top \theta^i}- \bm{\beta^i}- \bm{Y^i(x^{-i})x^i}= 0\\
 \quad -\bm{e}^\top \bm{g^i}-\bm{e}^\top \bm{\phi^i}\leq \frac{1}{\varepsilon_i}s, & -\bm{\tau^i - f^i} = \frac{1}{\varepsilon_i} \bm{m}\\
 \quad  -\bm{\tau^i + \phi^i} \leq \sigma_i \bm{m}, & \bm{\tau^i + \phi^i} \leq - \sigma_i \bm{m}\\
 \quad \bm{W\tau^i} \geq - \sigma_i \bm{h}, & \bm{Wf^i} \geq - \bm{h} \\
 \quad -\bm{f^i +g^i} \leq \bm{m}, & \bm{f^i +g^i} \leq - \bm{m} \\
 \quad  -\bm{\lambda^i} +\bm{\kappa^i} + \bm{W}^\top \bm{\xi^i}- \bm{\beta^i}= 0, \\
 \quad \bm{\lambda^i,\kappa^i, \delta^i, \nu^i, x^i} \geq \bm{0}, & \bm{\theta^i, \xi^i, \phi^i, g^i} \leq \bm{0} \\
 \quad \gamma \geq 0, & \\
  \end{array}}
 \end{equation}
where $\bm{Y^i(x^{-i})} \in \mathbb{R}^{(N \prod_{i=1}^N a_i)\times a_i}$ denotes the matrix such that 
 \begin{equation}
 \label{ooo} 
 vec(\bm{P})^\top \bm{Y^i(x^{-i})x^i} = \pi_i(\bm{P;x^{-i},x^i}),
  \end{equation} 
   parameter $\varepsilon_i $ denotes the risk level of player i and\\ $ \sigma_i = \frac{1-\varepsilon_i}{\varepsilon_i}$ is a fixed number $\forall i \in \{1,2,....N\}$.
\end{thm}
\section{\uppercase{The ambiguity set}}
\noindent In all DRG that we develop in this paper the ambiguity set have the following form:
\begin{equation}
 \label{mkm}
\Scale[0.7] {\mathcal{F} = \{ Q : Q[\bm{W} \cdot vec(\bm{\tilde{P}})\leq \bm{h}] = 1 , \;\; \mathbb{E}_Q [vec \bm{\tilde{P}}] = \bm{m}, \;\; \mathbb{E}_Q [\left \lVert vec (\bm{\tilde{P}}) - \bm{m} \right \rVert_1] \leq s \}}
\end{equation}
This, combined with different risk levels $\varepsilon_i$ for player $ i \in \{1,2,..N\}$ allows several variations of each distributionally robust game. By changing the values of ambiguity set's uncertain parameters $\bm{W,h, m}$ and $s$ and by assuming each time different risk attitudes for the players the set of DROE which constitute the solution of our problem can change dramatically.\\
At this point, we present the role of each uncertain parameter of the ambiguity set ~\eqref{mkm}.\\
Matrix $\bm{W} \in \mathbb{R}^{(m \times N \prod_{i=1}^N a_i)}$  and vector $\bm{h} \in \mathbb{R}^m $ are the two variables which represent the uncertainty polyhedral set in which the uncertain values of the payoff matrix should belong.
The maximum distance of all possible $vec(\bm{\tilde{P}})$ from the average vector $\bm{m}$ is denoted by scalar s. 
Finally, $\bm{m} \in \mathbb{R}^{N \prod_{i=1}^N a_i}$ is the vector that denotes the expected value of $vec(\bm{\tilde{P}})$ for each distribution that belongs to the ambiguity set. \\
\underline{Important assumption:} Vector $\bm{m}$ must belong to the bounded uncertainty polyhedral set of the payoff matrix $\bm{\tilde{P}}$. Otherwise the ambiguity set $\mathcal{F}$ will be empty.\\

\noindent\textbf{Special Cases of Distributionally Robust Games:}\\
Under certain conditions (special cases), the set of equilibria of distributionally robust finite game with ambiguity set like ~\eqref{mkm} is equivalent to that of a related finite game with complete payoff information (Nash Game) and with the same number of players and the same action spaces. Thus, when our game falls within these special cases one can simply solve the corresponding Nash Game.\\

The special cases of such games are studied in the next three propositions.
\begin{prop}
\label{lem1}
The set of equilibria of a distributionally robust game in which all players are risk neutral ($\varepsilon_i = 1$, $\forall i \in \{1,2,..N\}$) is equivalent to the set of equilibria of a Nash Game with fixed payoff matrix $\bm{\Psi}$ where $vec(\bm{\Psi}) = \bm{m}$. ($\bm{m}$ is the average vector of the ambiguity set ~\eqref{mkm}. ) 
 \end{prop}
 \begin{prop}
\label{lem2}
The set of equilibria of a distributionally robust game in which the parameter $s$ of the ambiguity set ~\eqref{mkm} is equal to zero(s=0) is equivalent to the set of equilibria of a Nash Game with fixed payoff matrix $\bm{M}$ where $vec(\bm{M})=\bm{m}$. ($\bm{m}$ is the average vector of the ambiguity set ~\eqref{mkm}. )
\end{prop}
\begin{prop}
\label{lem3}
The set of equilibria of a distributionally robust game that has as a support a single point is equivalent to the set of equilibria of a Nash Game with fixed payoff matrix the one that corresponds to this single point. Single point is named the unique payoff matrix which created from specific values of the matrix $\bm{W}$ and vector $\bm{h}$ of the ambiguity set. The values of matrix $\bm{W}$ and vector $\bm{h}$ are selected in order to make the uncertainty set $U= \{\bm{P}: \bm{W} \cdot vec(\bm{P})\leq \bm{h}\}$ singleton.
\end{prop}

\section{\uppercase{Illustrative example}}
\label{examplesdistribgames}
\noindent Having presented our distributionally robust games model we will now illustrate our approach with one concrete example.

\noindent \textbf{Distributionally Robust Inspection Game (DRIG)}\\
\emph{Problem Description:}\\
The DRIG is a two player game in which the row player is the employee (possible actions:Shirk or Work) and the column player is the employer(with possible actions Inspect or not Inspect). 
\begin{table}[H]
\caption{Normal Form Representation of DRIG}
\centering
\begin{tabular}{ |c|c|c| }
    \hline
      & Inspect & NotInspect \\ \hline
    Shirk & ($0 ,-\tilde{h}$) & (w, -w) \\ \hline
    Work & ($w-\tilde{g},\tilde{v}-w-\tilde{h}$) & ($w-\tilde{g},\tilde{v}-w$) \\
    \hline
  \end{tabular}
\label{robinspectiongame}
  \end{table}
The two players choose their actions simultaneously and then they receive the payoffs corresponding to the combination of their strategies. When the employee works he has cost $\tilde{g}$ and his employer has profit equal to $\tilde{v}$. Each inspection costs to the employer $\tilde{h}$ but if he inspects and finds the employee shirking then he does not pay him his wage w. In all other cases employee's wage is paid. All values except the payment w of the employee are uncertain. 

In the classical complete information Inspection game the parameters $\tilde{g}$, $\tilde{v}$ and $\tilde{h}$  are fixed ($\tilde{g}=\check{g}, \tilde{v}=\check{v}, \tilde{h}=\check{h}$) and in the robust approach $(\tilde{g},\tilde{v},\tilde{h}) \in [ \underline{g} , \overline{g} ]\times [ \underline{v} , \overline{v} ]\times [ \underline{h} , \overline{h} ]$ \cite{aghassi2006robust}.\\
In our new, distributionally robust approach, players have partial information about the probability distributions of the uncertain variables $\tilde{g},\tilde{v}$ and $\tilde{h}$ (about the probability distribution Q of the payoff matrix $\bm{\tilde{P}}$). In particular, the players do not know the exact distribution of the payoff matrix. They are only aware of a commonly known ambiguity set $\mathcal{F}$ of all possible probability distributions $Q$ that satisfy some specific properties. Subsequently, all players adopt a worst case CVaR approach to the uncertainty which is computed over all probability distributions within the set  $\mathcal{F}$. 
The introduction of the CVaR in the formulation of the game allows the two players to have different risk attitudes. Finally, the risk levels of the players are assumed to be common knowledge and none of the two players has private information.\\
For example, we may consider the DRIG in which the ambiguity set is given by:
\begin{equation}
  \label{Inspectionerequili}
\Scale[0.7] {{\mathcal{F}} = \{ Q : Q[(\tilde{g},\tilde{v},\tilde{h}) \in U] = 1 , \;\; \mathbb{E}_Q [vec (\tilde{\bm{P}})] = \bm{m}, \;\; \mathbb{E}_Q [\left \lVert vec (\tilde{\bm{P}}) - \bm{m} \right \rVert_1] \leq s \}}
\end{equation}
Where:$U = [ \underline{g} , \overline{g} ]\times [ \underline{v} , \overline{v} ]\times [ \underline{h} , \overline{h} ]$ ,  $s \geq 0$, and
\begin{equation}
 \tilde{\bm{P}} = \begin{pmatrix}
(0 ,-\tilde{h}) & (w, -w)  \\
(w-\tilde{g},\tilde{v}-w-\tilde{h})) & (w-\tilde{g},\tilde{v}-w) \end{pmatrix}
\end{equation}

\emph{Important assumption:} Vector $\bm{m}$ of the second constraint of the ambiguity set must belong to the bounded polyhedral uncertainty set of payoff matrix $\bm\tilde{P}$. For the DRIG, $\bm{m} \in  [ \underline{g} , \overline{g} ]\times [ \underline{v} , \overline{v} ]\times [ \underline{h} , \overline{h} ]$. Otherwise the ambiguity set will be empty.
\section{\uppercase{Numerical Experiment}}
\noindent In this section, we experimentally evaluate the new model of games described in this paper. In the interest of brevity, we only present one experiment. For more examples, experiments and detailed explanation of our computational method we refer the reader to chapter 5 of \cite{NLoizou}. \\

\noindent \textbf{The Experiment:}\\
\textbf{Fixed ambiguity set - Several Risk Levels}\\
\emph{What would happen to the number of equilibria and to the payments of the two players when the ambiguity set is kept fixed while the values of players' risk levels are varied.}\\

\noindent \emph{Computational Method:}\\
The method that we use to approximately compute the DROE and the players' payoffs at each equilibrium of any DRG is developed as follows:
\begin{enumerate}
\item Check if the ambiguity set of the DRG can be expressed like the general form of equation ~\eqref{mkm}. (The ambiguity set of DRIG has this property.)
\item Estimate the multi-linear system of equalities and inequalities whose dimension-reducing component-wise projection of the feasible solution set is equivalent with the set of equilibria of the DRG (see theorem ~\eqref{alldistribequilibria}) 
\item Find the feasible solutions of the multi-linear system and for each solution keep the components that correspond to the strategies of the players (projection of the solution). Additionally, compute the players' payoffs at each equilibrium. These are achieved using the YALMIP modelling language \cite{lofberg2004yalmip}, in Matlab 2014b.
\end{enumerate} 
All numerical evaluations of this chapter were conducted on a 2.27GHz, Intel Core i5 CPU 430 machine with 4GB of RAM.\\

\begin{figure}[h]
  \centering
  \subfigure[]{\includegraphics[scale=0.22]{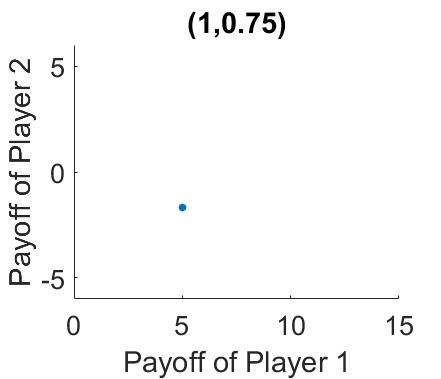}}\quad
  \subfigure[]{\includegraphics[scale=0.22]{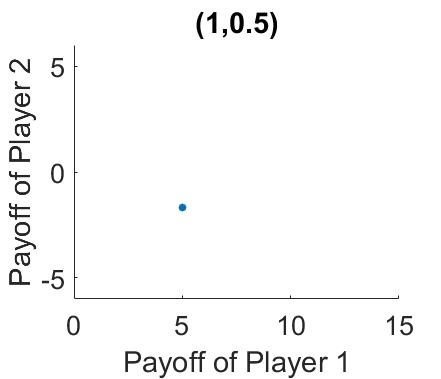}}
   \subfigure[]{\includegraphics[scale=0.22]{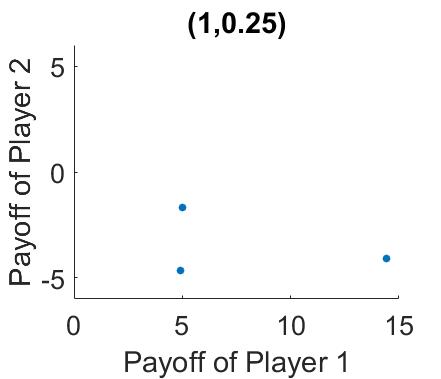}}\quad
  \subfigure[]{\includegraphics[scale=0.22]{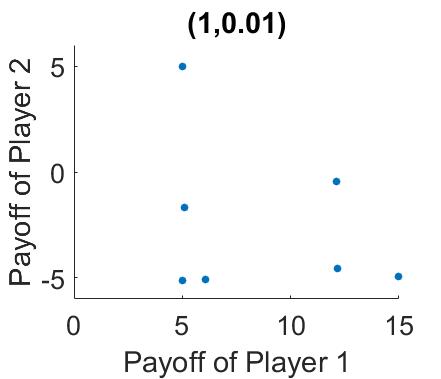}}
  \caption{{\fontsize{9}{10.8}\selectfont Graph representation of the payoffs of the two players at equilibria for different risk levels. Risk level of player 1 is kept fixed ($\varepsilon_1=1$) while player 2 has several levels of risk aversion. The title of each sub-figure denotes the risk levels of the two players: (Players 1's risk level, Players 2's risk level). }}
\label{fig:test1}
\end{figure}

More specifically, in the experiment of this section we assume that the uncertain parameters of the costs $(\tilde{g},\tilde{v},\tilde{h})$ must belong to $[8,12]\times [16,24]\times [4,6]$ and that the payment w of the employee is fixed at w=15. In addition, we assume that average vector $\bm{m}$ of the ambiguity set is the one that corresponds to the nominal\footnote{With ``nominal" we mean that the average vector takes the value of $vec(\bm{P})$ when the uncertain parameters $\tilde{g},\tilde{v}$ and $\tilde{h}$ are equal to the mid points of their intervals ($\tilde{g}=10, \tilde{v}=20, \tilde{h}=5$ and $w=15$). } version of the game $\bm{m}=\bm{m_1}=(0,-5,15,-15,5,0,5,5)^\top$ and without loss of generality that the maximum distance s of the third constraint of the ambiguity set is $s=4$.  \\
Therefore, since all variables of the ambiguity set are kept fixed \footnote{The matrix $\bm{W}$ and vector $\bm{h}$ of the first constraint of the ambiguity set are also fixed because the uncertain parameters of the payoff matrix  ($\tilde{g},\tilde{v}, \tilde{h}$) belong in a specific fixed uncertainty set} , we can derive that the ambiguity set of the form ~\eqref{mkm} is kept fixed in this experiment . The only variables that are allowed to change are the risk levels $\varepsilon_1$ and $\varepsilon_2$ of the two players.\\
The following tables and figures illustrate the number of equilibria and the players' payoffs at these equilibria for the aforementioned fixed ambiguity set while the players' risk levels change.\\
In particular Table ~\ref{tablkiuho} shows the equilibria of the previously described game when player 1  is risk neutral ($\varepsilon_1=1$) and player 2  has several risk attitudes. The players' payoffs at equilibria for each combination of the risk levels are given in Figure ~\ref{fig:test1}.\\

Notice that the first line of the Table ~\ref{tablkiuho} and Table ~\ref{risk2fixed} corresponds to the special case in which the two players of the game are risk neutral($\varepsilon_1=1, \varepsilon_2= 1$). This means that instead of solving the DRIG using the aforementioned computational method we can simply solve the corresponding Nash Game with fixed payoff matrix $\bm{M}$ where $vec{\bm{M}}= \bm{m}$: 
\begin{equation}
\tilde{\bm{M}} = \begin{pmatrix}
(0 ,-5) & (15, -15)  \\
(5,0) & (5,5) \end{pmatrix}
\end{equation}
This Complete Information Game has unique equilibrium equal to (1/3,2/3). That is, the employee(row player) shirks(plays action 1) with probability $x_1^1= 1/3$ and the employer(column player) inspects(plays action 1)with probability  $x_1^2= 2/3$.

\begin{table}[h]
\caption{DRIG: The equilibria for different values of Risk levels when the vector $\bm{m}$ of the ambiguity set take the nominal value and the maximum distance is $s=4$. Player 1 is risk neutral. His risk level is kept fixed ($\varepsilon_1=1$) while player 2 has several levels of risk aversion.}
\centering
    \scalebox{0.7}{\begin{tabular}{| l | p{5cm}  | } 
   \hline 
   Risk Levels & Equilibria  \\ 
   \hline 
    $\varepsilon_1=1, \varepsilon_2= 1$ & (1/3,2/3)  \\ 
   \hline 
   $\varepsilon_1=1, \varepsilon_2= 0.75$ & (0.333, 0.66) \\
    \hline 
    $\varepsilon_1=1, \varepsilon_2= 0.5$ & (0.333, 0.66)\\ 
    \hline 
    $\varepsilon_1=1, \varepsilon_2= 0.25$ & (0.333,0.66)(0.8179,0) (0.9342,0.7069)\\ 
    \hline 
      $\varepsilon_1=1, \varepsilon_2= 0.01$  &(1,0),(0,0.66), (1,0.66),(1,0.1941) (0.333,0.66), (0.9654,0.1387),(1,0.59)\\ 
    \hline 
   \end{tabular}}
\label{tablkiuho}
\end{table}

\begin{figure}[h]
  \centering
  \subfigure[]{\includegraphics[scale=0.22]{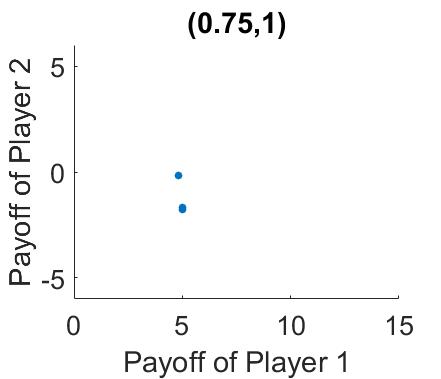}}\quad
  \subfigure[]{\includegraphics[scale=0.22]{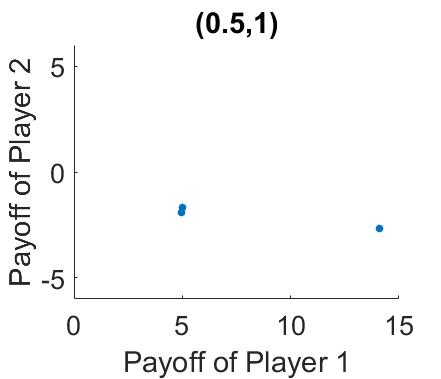}}
   \subfigure[]{\includegraphics[scale=0.22]{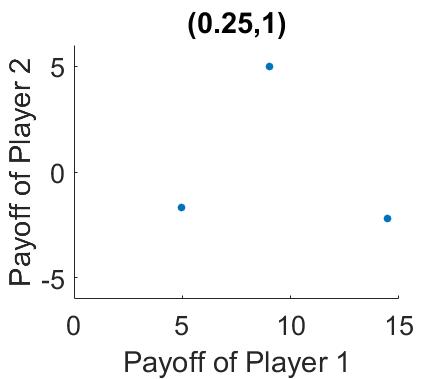}}\quad
  \subfigure[]{\includegraphics[scale=0.22]{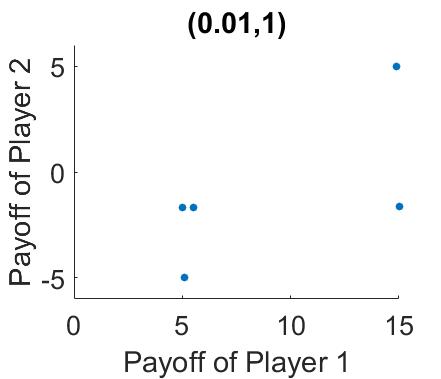}}
  \caption{{\fontsize{9}{10.8}\selectfont Graph representation of the payoffs of the two players at equilibria for different risk levels. Risk level of player 2 is kept fixed ($\varepsilon_2=1$) while player 1 has several levels of risk aversion. The title of each sub-figure denotes the risk levels of the two players: (Players 1's risk level, Players 2's risk level). }}
\label{fig:test2}
\end{figure}

Subsequently,  Table ~\ref{risk2fixed} shows the equilibria of the previously described game when player 2  is risk neutral ($\varepsilon_2=1$) and player 1  has several risk attitudes. The players' payoffs at equilibria for each combination of the risk levels are given in Figure ~\ref{fig:test2}.

\begin{table}[h]
\caption{DRIG: The equilibria for different values of Risk levels when the vector $\bm{m}$ of the ambiguity set take the nominal value and the maximum distance is $s=4$. Player 2 is risk neutral. His risk level kept fixed ($\varepsilon_2=1$) while player 1 has several levels of risk aversion.}
\centering
    \scalebox{0.7}{\begin{tabular}{| l | p{5cm} | } 
   \hline 
   Risk Levels & Equilibria  \\
   \hline 
    $\varepsilon_1=1, \varepsilon_2= 1$ & (1/3,2/3)  \\ 
   \hline 
   $\varepsilon_1=0.75, \varepsilon_2= 1$ & (0.333,0.666),(0.35,0.665), (0.2583,0.96)\\
    \hline 
    $\varepsilon_1=0.5, \varepsilon_2= 1$ & (0.333,0.666),(0.5379,0), (0.3842,0.66) \\ 
    \hline 
    $\varepsilon_1=0.25, \varepsilon_2= 1$ & (0.4427,0),(0.333,0.666), (0,0.3467) \\ 
    \hline 
      $\varepsilon_1=0.01, \varepsilon_2= 1$  & (0,0),(1,1), (0.333,0.666),(0.33,0), (0.335,1) \\ 
    \hline 
   \end{tabular}}
\label{risk2fixed}
\end{table}

\begin{figure}[h]
  \centering
  \subfigure[]{\includegraphics[scale=0.22]{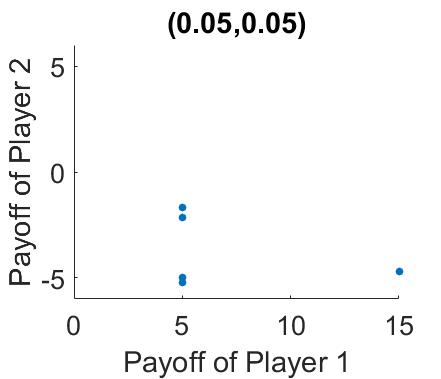}}\quad
  \subfigure[]{\includegraphics[scale=0.22]{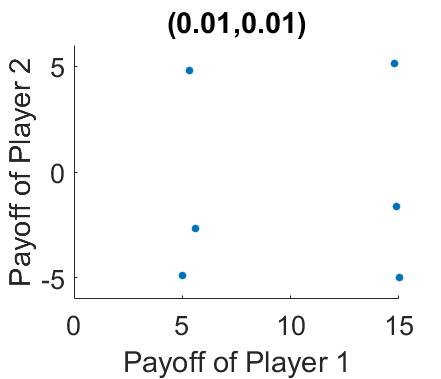}}
  \caption{{\fontsize{9}{10.8}\selectfont Graph representation of the players' payoffs at equilibria when their risk levels are $\varepsilon_1=\varepsilon_2=0.05$ and $\varepsilon_1=\varepsilon_2=0.01$.The title of each sub-figure denotes the risk levels of the two players: (Players 1's risk level,Players 2's risk level). }}
\label{fig:test}
\end{figure}

\textbf{Discussion of the Results:}\\
In standard optimization problems we know that as the decision maker becomes more risk averse his payoff always decreases. From the previous Figures ~\ref{fig:test1} and ~\ref{fig:test2} we can conclude that in game theory situation this is not always the case. We can not have a general rule,  since now the nature of the problem is more complicated. \\
In the DRG we assume that players are rational, so they can predict the outcome of the game and choose the strategies that form an equilibrium. For this reason, a difference at risk attitude of a player does not change only his decision but also the decisions of his opponents.\footnote{in DRG, risk attitude is assumed to be common knowledge}.
Hence, with increasing of risk aversion of one player the players' payoffs at equilibria may both increase or decrease depending the game. For example, in Figure ~\ref{fig:test1} at Subfigure (d) where the risk levels are (1,0.01) we can observe that for some equilibria the players have large payments and for some others very low.\\
To verify that in the DRIG we can not have a general rule about what happen in the payoffs of the two players when they choose to play strategies that form equilibria we also present Table ~\ref{005001} and Figure ~\ref{fig:test} witch illustrate the payoffs of the two players at equilibria when both of them are risk averse. More specifically when their risk levels are $\varepsilon_1=\varepsilon_2=0.05$ and $\varepsilon_1=\varepsilon_2=0.01$\\
\begin{table}[h]
\caption{DRIG: Equilibria when the players' risk levels are $\varepsilon_1=\varepsilon_2=0.05$ and $\varepsilon_1=\varepsilon_2=0.01$}
\centering
    \scalebox{0.7}{\begin{tabular}{| l | p{5cm} | } 
   \hline 
   Risk Levels & Equilibria  \\
   \hline 
    $\varepsilon_1=\varepsilon_2= 0.05$ & (1,0.66),(1,1)(0.95,0), (0.43,1),(0.333,0.666)  \\ 
    \hline 
      $\varepsilon_1=\varepsilon_2=0.01$  & (1,0),(0,0),(0.332,0), (0.5303,1),(1,0.78) \\ 
    \hline 
   \end{tabular}}
\label{005001}
\end{table}

\section{\uppercase{Conclusions}}
\noindent This paper combines Game Theory and Distributionally Robust Optimization to propose a novel model of incomplete information games without private information in which the players use distributionally robust optimization to cope with payoff uncertainty.\\
We showed that for specific ambiguity sets and risk levels, distributionally robust games constitute a true generalization of Nash games, Bayesian Games and Robust Games. Thus, any finite game of these three categories can be expressed as a distributionally robust game.\\
Subsequently, we proved that the set of equilibria of an arbitrary distributionally robust game with specified ambiguity set and without private information can be computed as the component-wise projection of the solution set of a multi-linear system of equations and inequalities. For special cases of such games we also showed equivalence to complete information finite games (Nash Games) with the same number of players and same action spaces.\\
Finally to concretize the idea of a distributionallly robust game we presented Distributionally Robust Inspection Game. We experimentally evaluated the new model of games and we studied how the number of equilibria and the players' payments change when the ambiguity set is fixed and the risk levels of the players are varied.\\
Our approach opens up many avenues for further development and research. For instance, the work of this paper can be generalized to the case of distributionally robust games involving potentially private information. Furthermore, interesting results might arise if we try to make similar work for more general classes of ambiguity sets.

\section*{\uppercase{Acknowledgements}}
\noindent The author is very grateful to Wolfram Wiesemann for helpful discussions and remarks. NL acknowledges support by the Leventis Foundation and Laura Wisewell Fund. Most of this work was done while at Imperial College London.

\bibliographystyle{apalike}
{\small
\bibliography{DRG}}

\section*{\uppercase{Appendix}}

\noindent  
\textbf{Proof of Theorem ~\ref{allequilibria}}:
Using equation ~\eqref{elnet4} we obtain the following results:\\
\underline{\emph{For Nash Games:}}
  If $ \mathcal{F} = \{ Q : \mathbb{E}_Q [\bm{\tilde{P}}] = \bm{\Psi} \}$ 
\begin{equation}
\Scale[0.7] {\begin{aligned}
\bm{x_i} \in \underset{u^i\in S_{a_i}} {\operatorname{argmax}} \,\inf_{Q \in \mathcal{F}} \; [\pi_i(\bm{\mathbb{E}_Q [\tilde{P}];x^{-i},u^i})] 
& = \underset{u^i\in S_{a_i}} {\operatorname{argmax}} \,\inf_{Q \in \mathcal{F}} \; [\pi_i(\bm{\Psi;x^{-i},u^i})]\\
&= \underset{u^i\in S_{a_i}} {\operatorname{argmax}} \;  [\pi_i(\bm{\Psi;x^{-i},u^i})]
\end{aligned}}
\end{equation}
which is equivalent to the formulation of Nash Equilibrium (see equation ~\eqref{lllll}).\\
\underline{\emph{For Bayesian Games:}}
If the ambiguity set is singleton, that is $ \mathcal{F} = \{ Q \}$.
\begin{equation}
  \Scale[0.7] { \bm{x_i} \in \underset{u^i\in S_{a_i}} {\operatorname{argmax}} \,\inf_{Q \in \mathcal{F}} \; \mathbb{E}_Q [\pi_i(\bm{\tilde{P};x^{-i},u^i})]  = \underset{u^i\in S_{a_i}} {\operatorname{argmax}} \,\mathbb{E}_Q [\pi_i(\bm{\tilde{P};x^{-i},u^i})]}
   \end{equation}\\  
which is equivalent to the formulation of Bayesian Nash Equilibrium (see equation ~\eqref{lmljsd}).\\  
\underline{\emph{For Robust Games:}}
If  $ \mathcal{F} = \{ Q : Q[\mathbb{E}_Q [\bm{\tilde{P}}] \in \mathcal{U}] = 1\} $ where $\mathcal{U} = \{ \bm{P} :\bm{W} \cdot vec(\bm{P})\leq \bm{h} \}$, then:
\begin{equation}
\Scale[0.7] {\bm{x_i} \in \underset{u^i\in S_{a_i}} {\operatorname{argmax}} \,\inf_{Q \in \mathcal{F}} \; [\pi_i(\bm{\mathbb{E}_Q [\tilde{P}];x^{-i},u^i})] = \underset{u^i\in S_{a_i}} {\operatorname{argmax}} \,\inf_{\bm{\tilde{S}} \in \mathcal{U}} \; [\pi_i(\bm{\tilde{S};x^{-i},u^i})] }
\end{equation}
where $\bm{\tilde{S}} =  \bm{\mathbb{E}_Q [\tilde{P}]};$\\
which is equivalent to the formulation of Robust Optimization Equilibrium Games (see equation ~\eqref{loioi}).\\

\noindent  
\textbf{Proof of Theorem ~\ref{alldistribequilibria}}:
By the Formulation of the DRG Condition 1 is equivalent to
\begin{equation} 
\label{sss}
\Scale[0.8] {x^i \in \underset{u^i\in S_{a_i}}{\operatorname{argmin}}\, \underset{Q \in \mathcal{F}}{\text{sup}}\; Q\text{-CVaR}_{\varepsilon_i} [-\pi_i(\bm{\tilde{P};x^{-i},u^i})]  \qquad  \forall i \in \{1,2,....,N\} }
\end{equation}
From \cite{rockafellar2000optimization} and \cite{rockafellar2002conditional} we know that :
\begin{equation} 
\Scale[0.7] {Q\text{-CVaR}_{\varepsilon_i} [-\pi_i(\bm{\tilde{P};x^{-i},u^i})] = \min_{\zeta_i \in \mathbb{R}} \; \zeta_i + \frac{1}{\varepsilon_i} \mathbb{E}_Q [ -\pi_i(\bm{\tilde{P};x^{-i},u^i}) - \zeta_i]^+ }
\end{equation}
where $[x]^+ = \max\{x,0\}$.\\
Therefore equation ~\eqref{sss} is equivalent to:
\begin{equation} 
 \Scale[0.7] {x^i \in \underset{u^i\in S_{a_i}}{\operatorname{argmin}}\,\underset{Q \in \mathcal{F}}{\text{sup}}\; \min_{\zeta_i \in \mathbb{R}} \; \zeta_i + \frac{1}{\varepsilon_i} \mathbb{E}_Q [ -\pi_i(\bm{\tilde{P};x^{-i},u^i}) - \zeta_i]^+ \quad \forall i \in \{1,..,N\}} 
\end{equation}
From Saddlepoint theorem ( Sion's minimax theorem \cite{sion1958general} )  we could exchange the order of supremum and infimum(minimum)\footnote{We can use Sion's minimax theorem since the function under consideration is convex-concave in its two arguments $Q \in \mathcal{F}$ and $\zeta_i \in \mathbb{R}$}   resulting: 
\begin{equation}
 \label{asaa}
 \Scale[0.7] {x^i \in \underset{u^i\in S_{a_i}}{\operatorname{argmin}}\, \min_{\zeta_i \in \mathbb{R}} \; \zeta_i + \frac{1}{\varepsilon_i}\,\underset{Q \in \mathcal{F}}{\text{sup}}\;\mathbb{E}_Q [ -\pi_i(\bm{\tilde{P};x^{-i},u^i}) - \zeta_i]^+ \quad \forall i \in \{1,..N\}} 
\end{equation}\\
From the moment problem theory, $ \, \underset{Q \in \mathcal{F}}{\text{sup}}\; \mathbb{E}_Q [ -\pi_i(\bm{\tilde{P};x^{-i},u^i}) - \zeta_i]^+ $ can be cast as the following problem: \\
\begin{equation}
\Scale[0.8] {\begin{array}{l@{\quad}l@{\qquad}l}
\displaystyle \text{maximize} & \displaystyle \int_{\mathcal{U}} [ -\pi_i(\bm{P;x^{-i},u^i}) - \zeta_i]^+  \mathrm{d}\mu ( vec (\bm{P})) \\
\displaystyle \text{subject to} & \displaystyle \mu \in \mathcal{M}_+ \mathbb{R}^{N \prod_{i=1}^N a_1} \\
& \displaystyle \int_{\mathcal{U}} \mathrm{d}\mu (vec(\bm{P})) = 1 \\
& \displaystyle \int_{\mathcal{U}} vec(\bm{P}) \mathrm{d}\mu (vec(\bm{P})) = \bm{m} \\
& \displaystyle \int_{\mathcal{U}} \left \lVert vec(\bm{P}) - \bm{m} \right \rVert_1 \mathrm{d}\mu (vec(\bm{P})) \leq s,
\end{array}}
\end{equation}
where $\mathcal{M}_+ \mathbb{R}^{N \prod_{i=1}^N a_i}$ is the set of non-negative measures supported on $\mathbb{R}^{N \prod_{i=1}^N a_i}$.\\
There is a duality theory for moment problems(see \cite{wiesemann2014distributionally} and \cite{natarajan2009constructing}) which implies that the following dual problem attains the same optimal value:
\begin{equation}
\Scale[0.7] {\begin{array}{l@{\quad}l@{\qquad}l}
\displaystyle \text{min} & \displaystyle \alpha_i + \bm{m}^\top \bm{\beta^i} + s \gamma_i \\
\displaystyle \text{s.t} & \displaystyle \alpha_i \in \mathbb{R}, \;\; \bm{\beta^i} \in \mathbb{R}^{N \prod_{i=1}^N a_i}, \;\; \gamma_i \in \mathbb{R}_+ \\
& \displaystyle \alpha_i + vec (\bm{\tilde{P}})^\top \bm{\beta^i} + \left \lVert vec (\bm{\tilde{P}}) - \bm{m} \right \rVert_1 \gamma_i \geq [ -\pi_i(\bm{\tilde{P};x^{-i},u^i}) - \zeta_i]^+.
\end{array}}
\end{equation}\\
Where the last inequality must be satisfied for all $\Scale[0.75] { \bm{P} \in \mathcal{U}}$.\\

By replacing the definition of $[\cdot]^+$:
\begin{equation}
\label{main}
\Scale[0.75] {\begin{array}{l@{\quad}l@{\qquad}l}
\displaystyle \text{min} & \displaystyle \alpha_i + \bm{m}^\top \bm{\beta^i} + s\gamma_i \\
\displaystyle \text{s.t} & \displaystyle \alpha_i \in \mathbb{R}, \;\; \bm{\beta^i} \in \mathbb{R}^{N \prod_{i=1}^N a_1}, \;\; \gamma_i \in \mathbb{R}_+ \\
& \displaystyle \alpha_i + vec (\bm{\tilde{P}})^\top \bm{\beta^i} + \left \lVert vec (\bm{\tilde{P}}) - \bm{m} \right \rVert_1 \gamma_i \geq  -\pi_i(\bm{\tilde{P};x^{-i},u^i}) - \zeta_i \\
& \displaystyle \alpha_i + vec (\bm{\tilde{P}})^\top \bm{\beta^i} + \left \lVert vec (\bm{\tilde{P}}) - \bm{m} \right \rVert_1 \gamma_i  \geq 0.
\end{array}}
\end{equation}
Where the last two inequalities must be satisfied for all $\Scale[0.75] { \bm{P} \in \mathcal{U}}$.\\

Substituting this dual formulation of $ \, \underset{Q \in \mathcal{F}}{\text{sup}}\; \mathbb{E}_Q [ -\pi_i(\bm{\tilde{P};x^{-i},u^i}) - \zeta_i]^+ $  into ~\eqref{asaa}, we obtain the following which we call Main Problem. The projection of the solution set of this problem will be the set of the equilibria that we desire.\\
\begin{equation}
\label{nana}
\Scale[0.7] {\begin{array}{l@{\quad}l@{\qquad}l}
\displaystyle \underset{u^i,\zeta_i,\alpha_i,\beta^i,\gamma_i}{\operatorname{min}} & \displaystyle \zeta_i + \frac{1}{\varepsilon_i}(\alpha_i + \bm{m}^\top \bm{\beta^i} + s \gamma_i) \\
\displaystyle \text{s.t} & \displaystyle \alpha_i \in \mathbb{R}, \;\; \bm{\beta^i} \in \mathbb{R}^{N \prod_{i=1}^N a_1}, \;\; \gamma_i \in \mathbb{R}_+, \zeta_i \in \mathbb{R} \\
& \displaystyle u^i \in S_{a_i}\\
& \displaystyle \alpha_i + vec (\bm{\tilde{P}})^\top \bm{\beta^i} + \left \lVert vec (\bm{\tilde{P}}) - \bm{m} \right \rVert_1 \gamma_i \geq -\pi_i(\bm{\tilde{P};x^{-i},u^i}) - \zeta_i \\
& \displaystyle \alpha_i + vec (\bm{\tilde{P}})^\top \bm{\beta^i} + \left \lVert vec (\bm{\tilde{P}}) - \bm{m} \right \rVert_1 \gamma_i  \geq 0.
\end{array}}
\end{equation}\\
Where the last two inequalities must be satisfied for all $\Scale[0.75] { \bm{P} \in \mathcal{U}}$.\\

This is now a `classical robust optimisation problem' and we use standard duality techniques to simplify the semi-infinite constraints.
We know that:$f(p) \geq k ,\,\, \forall p \in U \Leftrightarrow \min_{ p \in U} f(p) \geq k $ Therefore, using this the two robust constraints of the linear program ~\eqref{nana} become:
\begin{equation}
\label{dasi}
\Scale[0.7] {\underset{\bm{P} \in \mathcal{U}}{\operatorname{min}} [\alpha_i + vec (\bm{P})^\top \bm{\beta^i} + \left \lVert vec (\bm{P}) - \bm{m} \right \rVert_1 \gamma_i + \pi_i(\bm{P;x^{-i},u^i})] \geq -\zeta_i }
\end{equation}
and
\begin{equation}
\label{das}
\Scale[0.7] {\underset{\bm{P} \in \mathcal{U}}{\operatorname{min}} [\alpha_i + vec (\bm{P})^\top \bm{\beta^i} + \left \lVert vec (\bm{P}) - \bm{m} \right \rVert_1 \gamma_i]  \geq 0 } 
\end{equation}
The left hand sides of the constraints ~\eqref{dasi} and ~\eqref{das} are equivalent to the following problems ~\eqref{mlm1} and ~\eqref{mlm2} respectively. 
\begin{equation}
\label{mlm1}
\Scale[0.7] {\begin{array}{l@{\quad}l@{\qquad}l}
\displaystyle \underset{vec(\bm{P})}{\operatorname{min}} & \displaystyle \alpha_i + vec (\bm{P})^\top \bm{\beta^i} + \left \lVert vec (\bm{P}) - \bm{m} \right \rVert_1 \gamma_i + \pi_i(\bm{P;x^{-i},u^i})\\
\displaystyle \text{subject to} & \displaystyle \bm{W} \cdot vec(\bm{P})\leq \bm{h}.
\end{array}}
\end{equation}
and
\begin{equation}
\label{mlm2}
\Scale[0.7] {\begin{array}{l@{\quad}l@{\qquad}l}
\displaystyle \underset{vec(\bm{P})}{\operatorname{min}} & \displaystyle \alpha_i + vec (\bm{P})^\top \bm{\beta^i} + \left \lVert vec (\bm{P}) - \bm{m} \right \rVert_1 \gamma_i \\
\displaystyle \text{subject to} & \displaystyle \bm{W} \cdot vec(\bm{P})\leq \bm{h}.
\end{array}}
\end{equation}
In turn these programs are equivalent to:
\begin{equation}
\label{kti1}
\Scale[0.7] {\begin{array}{l@{\quad}l@{\qquad}l}
\displaystyle \underset{vec(\bm{P}),\eta}{\operatorname{min}} & \displaystyle \alpha_i + vec (\bm{P})^\top \bm{\beta^i} + \gamma_i \sum\limits_{j=1}^{N\prod_{i=1}^N a_1} \eta_j +  vec(\bm{P})^\top \bm{Y^i(x^{-i})u^i}\\
\displaystyle \text{s.t} & \displaystyle \bm{W} \cdot vec(\bm{P})\leq \bm{h}\\
& \displaystyle \eta_j \geq {vec(\bm{P})}_j - \bm{m}_j \qquad \forall j= 1,2,...N\prod_{i=1}^N a_i& \displaystyle \\
& \displaystyle \eta_j \geq \bm{m}_j - {vec(\bm{P})}_j \qquad \forall j= 1,2,...N\prod_{i=1}^N a_i.& \displaystyle 
\end{array}}
\end{equation}
and 
\begin{equation}
\label{kti2}
\Scale[0.7] {\begin{array}{l@{\quad}l@{\qquad}l}
\displaystyle \underset{vec(\bm{P}),\eta}{\operatorname{min}} & \displaystyle \alpha_i + vec (\bm{P})^\top \bm{\beta^i} + \gamma_i \sum\limits_{j=1}^{N\prod_{i=1}^N a_i} \eta_j \\
\displaystyle \text{subject to} & \displaystyle \bm{W} \cdot vec(\bm{P})\leq \bm{h}\\
& \displaystyle \eta_j \geq {vec(\bm{P})}_j - \bm{m}_j & \displaystyle \forall j= 1,2,...N\prod_{i=1}^N a_i\\
& \displaystyle \eta_j \geq \bm{m}_j - {vec(\bm{P})}_j & \displaystyle \forall j= 1,2,...N\prod_{i=1}^N a_i.
\end{array}}
\end{equation}
where $\eta_j= |{vec(\bm{P})}_j - \bm{m}_j|,\,\, \forall j= 1,2,...N\prod_{i=1}^N a_i$ and $\bm{Y^i(x^{-i})}$ is as defined in ~\eqref{ooo}.\\
The dual problems of ~\eqref{kti1} and ~\eqref{kti2} are respectively the following:
\begin{equation}
\label{dual1}
\Scale[0.7] {\begin{array}{l@{\quad}l@{\qquad}l}
\displaystyle \underset{\delta^i,\beta^i,\theta^i}{\operatorname{max}} & \displaystyle \alpha_i - \bm{m^\top \delta^i} +\bm{m}^\top \bm{\nu^i} + \bm{h^\top \theta^i}\\
& \displaystyle -\bm{\delta^i} +\bm{\nu^i} + \bm{W}^\top \bm{\theta^i}- \bm{\beta^i}-\bm{Y^i(x^{-i})u^i}= 0\\
& \displaystyle \bm{\delta^i} +\bm{\nu^i} - \gamma_i\bm{e} \leq \bm{0}\\
& \displaystyle \bm{\delta^i} \geq \bm{0}, \;\; \bm{\nu^i} \geq \bm{0}, \;\; \bm{\theta^i} \leq \bm{0}.
\end{array}}
\end{equation}
\begin{equation}
\label{dual2}
\Scale[0.7] {\begin{array}{l@{\quad}l@{\qquad}l}
\displaystyle \underset{\lambda^i,\kappa^i,\xi^i}{\operatorname{max}} & \displaystyle \alpha_i - \bm{m^\top \lambda^i} +\bm{m}^\top \bm{\kappa^i} + \bm{h^\top \xi^i}\\
& \displaystyle -\bm{\lambda^i} +\bm{\kappa^i} + \bm{W}^\top \bm{\xi^i}- \bm{\beta^i}= 0\\
& \displaystyle \bm{\lambda^i} +\bm{\kappa^i} - \gamma_i\bm{e} \leq \bm{0}\\
& \displaystyle \bm{\lambda^i} \geq \bm{0}, \;\; \bm{\kappa^i} \geq \bm{0}, \;\; \bm{\xi^i} \leq \bm{0}.
\end{array}}
\end{equation}
We know that:$ \exists p \in U: f(p) \geq K \Leftrightarrow \max_{p\in U} f(p) \geq K$. Subsequently, we substitute the last two problems ~\eqref{dual1} and ~\eqref{dual2} in the Main Problem ~\eqref{nana}. Therefore for each player $i \in\{1,2,...N\},\,\, \exists \,\,\, \alpha_i,\gamma_i, \zeta_i \in \mathbb{R}, \bm{\beta^i,\lambda^i, \kappa^i,\delta^i, \nu^i} \in \mathbb{R}^{N \prod_{i=1}^N a_i}$ and $\bm{\xi^i, \theta^i} \in \mathbb{R}^m $ such that $ (\bm{x^i,\beta^i,\lambda^i, \kappa^i,\delta^i, \nu^i\xi^i, \theta^i,}\alpha_i, \gamma_i, \zeta_i)$ is a minimizer of:
\begin{equation}
\Scale[0.7] {\begin{array}{l@{\quad}l@{\qquad}l}
\displaystyle \underset{u^i,\alpha_i,\beta^i,\gamma_i, \zeta_i, \lambda^i, \kappa^i,\xi^i,\delta^i,\nu^i,\theta^i}{\operatorname{min}} & \displaystyle \zeta_i +  \frac{1}{\varepsilon_i}\alpha_i + \frac{1}{\varepsilon_i}\bm{m}^\top \bm{\beta^i} + \frac{1}{\varepsilon_i}s\gamma_i\\
& \displaystyle \bm{e}^\top \bm{u^i}=1\\
& \displaystyle \alpha_i - \bm{m^\top \lambda^i} +\bm{m}^\top \bm{\kappa^i} + \bm{h^\top \xi^i} \geq 0\\
& \displaystyle -\bm{\lambda^i} +\bm{\kappa^i} + \bm{W}^\top \bm{\xi^i}- \bm{\beta^i}= 0\\
& \displaystyle \bm{\lambda^i} +\bm{\kappa^i} - \gamma_i\bm{e} \leq \bm{0}\\
& \displaystyle \alpha_i - \bm{m^\top \delta^i} +\bm{m}^\top \bm{\nu^i} + \bm{h^\top \theta^i} + \zeta_i \geq 0\\
& \displaystyle -\bm{\delta^i} +\bm{\nu^i} + \bm{W}^\top \bm{\theta^i}- \bm{\beta^i}-\bm{Y^i(x^{-i})u^i}= 0\\
& \displaystyle \bm{\delta^i} +\bm{\nu^i} - \gamma_i\bm{e} \leq \bm{0}\\
& \displaystyle \bm{\lambda^i} \geq \bm{0}, \;\; \bm{\kappa^i} \geq \bm{0}, \;\; \bm{\xi^i} \leq \bm{0}\\
& \displaystyle \bm{\delta^i} \geq \bm{0}, \;\; \bm{\nu^i} \geq \bm{0}, \;\; \bm{\theta^i} \leq \bm{0}\\
& \displaystyle \bm{u^i} \geq \bm{0}, \;\; \gamma_i \geq 0.
\end{array}}
\end{equation}
whose dual is:
\begin{equation}
\Scale[0.7] {\begin{array}{l@{\quad}l@{\qquad}l}
\displaystyle \underset{\tau^i,\rho_i,f^i,\phi^i,g^i,}{\operatorname{max}} & \displaystyle \rho_i\\
& \displaystyle-\bm{e}^\top \bm{g^i}-\bm{e}^\top \bm{\phi^i}\leq \frac{1}{\varepsilon_i}s \\
& \displaystyle-\bm{\tau^i - f^i} = \frac{1}{\varepsilon_i} \bm{m} \\
& \displaystyle -\bm{\tau^i + \phi^i} \leq \sigma_i \bm{m} \\
& \displaystyle\bm{\tau^i + \phi^i} \leq - \sigma_i \bm{m} \\
& \displaystyle\bm{W\tau^i} \geq - \sigma_i \bm{h} \\
& \displaystyle -\bm{f^i +g^i} \leq \bm{m} \\
& \displaystyle \bm{f^i +g^i} \leq - \bm{m} \\
& \displaystyle \bm{Wf^i} \geq - \bm{h} \\ 
& \displaystyle \rho_i\bm{e}^\top \leq \bm{f^\top Y^i(x^{-i})} \\ 
& \displaystyle \bm{\phi^i} \leq \bm{0}, \;\;  \bm{g^i} \leq 0.
\end{array}}
\end{equation} 
Condition 2 follows from strong linear programming duality.
The reverse direction (Condition 2 $\Longrightarrow$ Condition 1) is also holds as all steps of our proof are based on the equivalence of the two parts.\\

\noindent  
\textbf{Proof of Proposition ~\ref{lem1}}: 
The second constraint of the ambiguity set $\mathcal{F}$ is  $\mathbb{E}_Q [vec \bm{\tilde{P}}] = \bm{m}$. Therefore if we denote with $\bm{\Psi}$ the matrix for which $vec(\bm{\Psi}) = m$ then $\bm{\Psi}=\mathbb{E}_Q [ \bm{\tilde{P}}]$ and with use of equation ~\eqref{elnet4}:
\begin{equation}
\Scale[0.7] {\begin{aligned}
 \bm{x_i} \in \underset{u^i\in S_{a_i}} {\operatorname{argmax}} \,\inf_{Q \in \mathcal{F}} \; [\pi_i(\bm{\mathbb{E}_Q [\tilde{P}];x^{-i},u^i})]  
& = \underset{u^i\in S_{a_i}} {\operatorname{argmax}} \,\inf_{Q \in \mathcal{F}} \; [\pi_i(\bm{\Psi;x^{-i},u^i})] \\
& = \underset{u^i\in S_{a_i}} {\operatorname{argmax}}  [\pi_i(\bm{\Psi;x^{-i},u^i})] 
\end{aligned}}
\end{equation}
which is equivalent to the formulation of Nash Equilibrium (see equation ~\eqref{lllll}).\\

\noindent  
\textbf{Proof of Proposition ~\ref{lem2}}: For $s=0$ the third constraint of the ambiguity set ~\eqref{mkm} becomes: $
\mathbb{E}_Q [\left \lVert vec (\bm{\tilde{P}}) - \bm{m} \right \rVert_1] \leq 0 $.
Then, since all values inside the expectation operator $\mathbb{E}_Q $ are positive we have that $
\mathbb{E}_Q [\left \lVert vec (\bm{\tilde{P}}) - \bm{m} \right \rVert_1] = 0$ and that $
Q [\left \lVert vec (\bm{\tilde{P}}) - \bm{m} \right \rVert_1 = 0] =1 
$ which is equivalent to 
\begin{equation}
Q [ vec (\bm{\tilde{P}_i}) - \bm{m_i} = 0] =1 ,\,\,\, \forall i \in \{1,2,... \mathbb{R}^{N \prod_{i=1}^N a_1} \}.\\
\end{equation}
Therefore if $s\longrightarrow0$ the third constraint of the ambiguity set is equivalent to: $Q [ vec (\bm{\tilde{P}}) - \bm{m} = 0] =1$
which means that $vec (\bm{\tilde{P}}) = \bm{m}$ for all distributions of the ambiguity set. Therefore:
\begin{equation}
\Scale[0.7] { \begin{aligned}
\mathcal{F} 
& = \{ Q : Q[\bm{W} \cdot vec(\bm{\tilde{P}})\leq \bm{h}] = 1 , \;\; \mathbb{E}_Q [vec \bm{\tilde{P}}] = \bm{m}, \;\;  Q[vec (\bm{\tilde{P}}) = \bm{m}] = 1 \}\\
& =  \{ Q : Q[vec (\bm{\tilde{P}}) = \bm{m}] = 1 \}
\end{aligned}}
\end{equation}
Using the definition of Nash Equilibrium we can find now the equivalence between our game and a Nash Game.
\begin{equation}
\Scale[0.7] {\begin{aligned}
\bm{x_i} \in \underset{u^i\in S_{a_i}}{\operatorname{argmin}}\,\underset{Q \in \mathcal{F}}{\text{sup}}\; Q\text{-CVaR}_{\varepsilon_i} [-\pi_i(\bm{\tilde{P};x^{-i},u^i})]
& = \underset{u^i\in S_{a_i}}{\operatorname{argmin}}\, [-\pi_i(\bm{M;x^{-i},u^i})] \\
& = \underset{u^i\in S_{a_i}} {\operatorname{argmax}}  [\pi_i(\bm{M;x^{-i},u^i})] 
\end{aligned}}
\end{equation}
where $vec(\bm{M})=\bm{m}$.\\

\noindent  
\textbf{Proof of Proposition ~\ref{lem3}}:  The uncertainty set $U$ is a singleton. Therefore the first constraint of the ambiguity set ~\eqref{mkm} is equivalent to: $Q[\bm{\tilde{P}} = \bm{C} ] = 1 $
where $\bm{C}$ denotes the support single point, the only matrix of set $U$. Thus the ambiguity set becomes:
\begin{equation}
\Scale[0.8] {\mathcal{F} = \{ Q : Q[\bm{\tilde{P}} = \bm{C} ] = 1  , \;\; \mathbb{E}_Q [vec \bm{\tilde{P}}] = \bm{m}, \;\;  Q[vec (\bm{\tilde{P}}) = \bm{m}] = 1 \}}\\
\end{equation}
where $vec(\bm{C})= m$.\\
Using this, the desired result follows:\\
\begin{equation}
\Scale[0.7] {\begin{aligned}
\underset{u^i\in S_{a_i}}{\operatorname{argmin}}\,\underset{Q \in \mathcal{F}}{\text{sup}}\; Q\text{-CVaR}_{\varepsilon_i} [-\pi_i(\bm{\tilde{P};x^{-i},u^i})]
& = \underset{u^i\in S_{a_i}}{\operatorname{argmin}}\,Q\text{-CVaR}_{\varepsilon_i} [-\pi_i(\bm{C;x^{-i},u^i})] \\
& = \underset{u^i\in S_{a_i}}{\operatorname{argmin}}\, [-\pi_i(\bm{C;x^{-i},u^i})] \\
& = \underset{u^i\in S_{a_i}} {\operatorname{argmax}}  [\pi_i(\bm{C;x^{-i},u^i})] 
\end{aligned}}
\end{equation}

\vfill
\end{document}